\documentclass[prl,aps,onecolumn,superscriptaddress,showpacs]{revtex4}
\usepackage{amsfonts}

\usepackage{dcolumn}
\usepackage{bm}
\usepackage{tikz}
\usepackage{color}
\usepackage[colorlinks=true,citecolor=blue,urlcolor=blue]{hyperref}
\usepackage{amsmath,amsfonts,amssymb,times,natbib}
\usepackage[standard]{ntheorem}

\newcommand{\ket}[1]{\ensuremath{| #1 \rangle}}

\begin{document}
\title{Non-adiabatic holonomic quantum computation in linear system-bath coupling}

\author{Chunfang Sun}
\affiliation{School of Physics, Northeast Normal University, Changchun 130024, People's Republic of China}

\author{Gangcheng Wang}
\affiliation{School of Physics, Northeast Normal University, Changchun 130024, People's Republic of China}

\author{Chunfeng Wu}
\affiliation{Pillar of Engineering Product Development, Singapore University
of Technology and Design, 8 Somapah Road, Singapore 487372}
\author{Haodi Liu}
\affiliation{School of Physics, Northeast Normal University, Changchun 130024, People's Republic of China}

\author{Xun-Li Feng}
\affiliation{Department of Physcis, Shanghai Normal University, Shanghai 200234, People's Republic of China}

\author{Jing-Ling Chen}
\affiliation{Theoretical Physics Division, Chern Institute of
Mathematics, Nankai University, Tianjin 300071, People's Republic of China}
\affiliation{Centre for Quantum Technologies, National University of Singapore, 3 Science Drive 2, Singapore 117543}

\author{Kang Xue}
\affiliation{School of Physics, Northeast Normal University, Changchun 130024, People's Republic of China}

\date{\today}

\pacs{03.65.Ud,
03.67.Mn,
42.50.Xa}

\maketitle

\textbf{Non-adiabatic holonomic quantum computation in decoherence-free subspaces protects quantum information from control imprecisions and decoherence.  For the non-collective decoherence that each qubit has its own bath, we show the implementations of two non-commutable holonomic single-qubit gates and one holonomic nontrivial two-qubit gate that compose a universal set of non-adiabatic holonomic quantum gates in decoherence-free-subspaces of the decoupling group, with an encoding rate of $\frac{N-2}{N}$. The proposed scheme is robust against control imprecisions and the non-collective decoherence, and its non-adiabatic property ensures less operation time. We demonstrate that our proposed scheme can be realized by utilizing only two-qubit interactions rather than many-qubit interactions. Our results reduce the complexity of practical implementation of holonomic quantum computation in experiments. We also discuss the physical implementation of our scheme in coupled microcavities.
}

\vspace{3mm}

Holonomic quantum computation (HQC), first proposed by Zanardi and Rasetti \cite{Zanardi8},is a general procedure for implementing quantum gates using non-Abelian geometric phases. In HQC, unitary operations can be implemented by varying the system Hamiltonian with degenerate energy levels to make the system evolve along a closed path in the parameter space. The unitary operations are determined only by the shape of the closed path, not on the details of the evolution. The property of HQC against control imprecisions leads to robust quantum operations. Thus HQC has become one promising quantum computation paradigm and attracted more and more interests recently \cite{Jones,Duan,Zhu1,Wu,Cen,Zhang3,Feng,Tong,Xu,Xu2,Abdumalikov,Feng1,Zhang2}.
The initial HQC is based on adiabatic evolution requiring long evolution time for the desired parametric control. To deal with this drawback, non-adiabatic HQC based on non-adiabatic non-Abelian geometric phases \cite{Anandan} has been proposed in Ref. \cite{Tong} and experimentally demonstrated in \cite{Abdumalikov,Feng1}.

Apart from errors in the control process, decoherence often caused by unavoidable interaction with environment is another main practical obstacle in quantum information processing (QIP). Various methods have been presented to protect quantum information against decoherence, such as symmetry-aided passive decoherence-free subspaces (DFSs) \cite{Lidar8} and noiseless subsystems (NSs) \cite{Knill} approaches, as well as active dynamical decoupling (DD) \cite{Viola} techniques.
The basic idea of DFSs and NSs is to utilize the natural symmetry of the system-environment interaction. Information stored in subspace spanned by the quantum states or subsystems are unaffected by the interaction with the environment. DFSs and NSs have been explored extensively in various physical systems  \cite{Kwiat,Kielpinski,Ollerenshaw,Mohseni,Bourennane,Zanardi5,Zanardi6}.
DD \cite{Viola} tackles decoherence by suppressing the system-environment interaction through stroboscopic pulsing of the system and it is thus called active approach against decoherence. As shown in the literatures \cite{Viola9,Khodjasteh,Yang,Viola1,Lidar7,West1}, DD not only can be used to preserve arbitrary state in quantum memories, it is also compatible with gate operations used for QIP in principle, essentially by designing DD operations that commute with the gate operations. Experimental demonstrations of DD protecting quantum gates have been recently  achieved in different physical systems \cite{Barthel,van}. Therefore, if the system-environment interaction has naturally available symmetries, one can use DFSs/NSs to encode and store quantum information. However, often times in practical applications such symmetries are imperfect, and hence  DFSs/NSs itself is not enough for protecting quantum information. In this case the combination of the active DD and the passive DFSs/NSs offers effective method to mitigate the negative effect of decoherence  \cite{Byrd,Wu1,West1,Viola9}.

To protect quantum information from both control imprecisions and the detrimental effects of the environment, the schemes hybridizing HQC with DFSs based on adiabatic evolution have been proposed \cite{Wu,Cen,Zhang3}. In order to avoid the long run time required by adiabatic evolution,  Refs. \cite{Xu,Xu2} has shown that non-adiabatic HQC can be realized in DFSs that are insensitive to the collective dephasing errors. For the general errors that each qubit has its own bath, the implementation of non-adiabatic holonomic gates can be protected from decoherence by resorting to the DD approach. According to the DD, undesirable couplings between system and environment can be effectively averaged out by utilizing repetition of fast external control operations. Due to the requirement of fast pulses, DD provides relatively less resource-demand protection for quantum information. However, the non-adiabatic HQC together with the integration of DD and DFSs/NSs has not been well explored. Very recently, Xu and Long~\cite{Long14} proposed a non-adiabatic HQC scheme based on two-qubit interactions and the scheme is robust against non-collective decoherence, by encoding three physical qubits to one logical qubit. Consider the scalability of the proposed quantum gates to many logical qubits, the scheme~\cite{Long14} requires a lot of resource. Thus more easily achievable scheme with a better encoding rate and against control imprecisions as well as non-collective decoherence is of great significance from the experimental perspective. In this work we address the issue by presenting a non-adiabatic HQC scheme against non-collective decoherence. We consider a linear system-bath interaction Hamiltonian in which each qubit has its own bath and provide a universal set of nonadiabatic holonomic quantum gates by presenting two noncommuting single-logical-qubit gates and one nontrivial two-logical-qubit gate in DFSs of a decoupling group. The encoding strategy used here is to encode $N$ physical qubits to $(N-2)$ logical qubits, and hence our scheme largely reduces the complexity of experiments.

\vspace{8pt}
\noindent{\bf Results}

We first recall the active DD technique \cite{Viola,Viola1} which is to be used to suppress the system-bath interaction later. In general, the interaction Hamiltonian without DD is of the form,
$H_{SB}=\sum_{\alpha} S_{\alpha}\otimes B_{\alpha}$,
where each $S_{\alpha}$  and $B_{\alpha}$ are pure-system operator and pure-bath operator, respectively. To suppress error, consider a group $\mathcal{G}\equiv \{g_{j}\}$, $j=0,1,...,|\mathcal{G}|-1$, of unitary
transformations $g_{j}$ acting purely on the system with $g_{0}\equiv\openone$ and $|\mathcal{G}|\equiv$ order ($\mathcal{G}$)
denoting the number of group elements. Assuming that each such pulse $g_{j}$ is effectively instantaneous and their temporal separation is $\Delta t$,  a full cycle time is $T_{c}=|\mathcal{G}|\Delta t$, and the natural propagator is $U_{0}(\Delta t)=\exp(-i H \Delta t)$. Then the evolution of the whole system with DD over a single cycle time is given by
$U(T_{c})=\prod_{j=0}^{|\mathcal{G}|-1}g_{j}^{\dagger}U_{0}(\Delta t)g_{j}\equiv e^{-iH_{\rm eff}T_{c}}$,
where $H_{\rm eff}$ denotes the resulting effective Hamiltonian. In the ideal limit
of arbitrarily fast control $T_{c}\rightarrow 0$, $H_{\rm eff}$ approaches
$H \longmapsto H_{\rm eff}=\frac{1}{|\mathcal{G}|}\sum_{g_{j}\in \mathcal{G}}g_{j}^{\dagger}H g_{j}\equiv \Pi_{\mathcal{G}(H)}$.
Note that $[H_{\rm eff},g_{j}]=0$ for $\forall g_{j}\in \mathcal{G}$,
thereby the decoupled evolution is symmetrized according to $\mathcal{G}$.

A decomposition of the system Hilbert space $\mathcal{H}_{S}$ can be induced by
the decoupling group $\mathcal{G}$  via its group algebra $\mathbb{C}\mathcal{G}$ and its
commutant algebra $\mathbb{C}\mathcal{G}^{'}$ as follows \cite{Viola1,Zanardi5}:
$\mathcal{H}_{S}  \cong \oplus_{J} \mathbb{C}^{n_{J}}\otimes \mathbb{C}^{d_{J}}$, 
$\mathbb{C}\mathcal{G}\cong \oplus_{J} \openone_{n_{J}}\otimes M_{d_{J}}$, and $\mathbb{C}\mathcal{G}^{'}= \oplus_{J} M_{n_{J}}\otimes \openone_{d_{J}}$.
Here the $J$-th irreducible representation (irrep), with the dimension $d_{J}$ , appears with the multiplicity $n_{J}$, while $M_{d}$ and $\openone_{d}$  are, respectively, the complex-valued $d\times d$   matrices and the $d\times d$ identity matrix.
We   encode the computational state into the left factor $\mathbb{C}^{n_{J}}$, the effective Hamiltonian $H_{\rm eff}$ needs to act trivially on $\mathbb{C}^{n_{J}}$. A necessary and sufficient condition is
$H_{\rm eff}\cong \oplus_{J} \lambda_{J}\openone_{n_{J}}\otimes \openone_{d_{J}}$ $(\lambda_{J}\in \mathbb{C})$.
In this case subsystems $\mathbb{C}^{n_{J}}$ are called NSs. When $d_{J}=1$, the DFSs case arises.

We consider a linear system-bath interaction Hamiltonian which is described by,
\begin{eqnarray}\label{HSB}
H_{SB}=\sum_{\alpha=x,y,z}\sum_{i}\sigma_{i}^{\alpha}\otimes B_{i}^{\alpha},
\end{eqnarray}
where $\sigma_{i}^{\alpha}$ are Pauli matrices acting on the $i$-th qubit and $B_{i}^{\alpha}$ are  arbitrary bath operators.
In this noise model, each qubit has its own bath. The decoupling group for $N$-qubit can be selected as \cite{Viola1}:
$\mathcal{G}=\{\openone^{\otimes N},X^{\otimes N},Y^{\otimes N},Z^{\otimes N}\}$,
where the pulses $X=\sigma_{x}$, $Z=\sigma_{z}$ and $Y=ZX=i\sigma_{y}$.
 Based on  $H_{\rm eff}$, the resulting average system-bath interaction becomes $H_{SB}^{'}=0$, which implies that the system is decoupled from the bath up to first-order at the time instant $t=T_{c}$.

Suppose that $N$ is even, $\mathcal{G}$ is an Abelian group with order $|\mathcal{G}|=4$, thus all the irreps of $\mathcal{G}$ are 1-dimensional (i.e., $d_{J}=1$), and the number of irreps is the order of the group. The group algebra $\mathbb{C}\mathcal{G}$ can be written as
$\mathbb{C}\mathcal{G}=\bigoplus_{J=1}^{4}c_{J}\openone_{2^{(N-2)}}$,
where $n_{J}=2^{(N-2)}$. Therefore each of the four equivalent subspaces (DFSs) is able to encode $(N-2)$ logical qubits to make universal quantum computation.
For instance, the $\mathcal{G}$-invariant subspace $\lambda=\{1,1,1,1\}$, representing a set of eigenvalues of decoupling group elements, is spanned
by the N-qubit quantum states $(|r\rangle+|\mathrm{NOT}(r)\rangle)/\sqrt{2}$, with $r$ containing an even number of $1's$ of length $N$.

For the system-bath interaction form (\ref{HSB}), the decoupling group $\mathcal{G}$ used to decouple the system from the bath up to first-order at the time instant $t=T_{c}$, has four equivalent $2^{(N-2)}$-dimensional DFSs with $N$ being even. Each of the four equivalent DFSs is able to encode $(N-2)$ logical qubits to make universal quantum computation \cite{Viola1} (i.e.,  there are $(N-2)$ logical qubits in each DFS that will be unaffected by the system-bath interaction). In the following, we utilize one of the four equivalent $\mathcal{G}$-invariant DFSs (i.e., $\lambda=\{1,1,1,1\}$) to encode our qubits.
The $(N-2)$ logical qubits are encoded in such subspace and the logical states are
 \begin{eqnarray}\label{logical states}
|r_{1}\rangle_{L}=\frac{1}{\sqrt{2}}(|0\rangle |r_{1}\rangle |0\rangle+|1\rangle|\mathrm{NOT}(r_{1})\rangle|1\rangle),\nonumber \\
|r_{2}\rangle_{L}=\frac{1}{\sqrt{2}}(|1\rangle |r_{2}\rangle|0\rangle+|0\rangle|\mathrm{NOT}(r_{2})\rangle|1\rangle),
\end{eqnarray}
where $|r_{1}\rangle_{L}$ and $|r_{2}\rangle_{L}$ are the logical states of $(N-2)$ logical qubits and the subscript L is used to denote that the states (or the operators) are logical states (or operators). $|r_{1}\rangle$ and $|r_{2}\rangle$ are the quantum states of  $(N-2)$ physical qubits from the $2$-th to the $(N-1)$-th physical qubits,  with $r_{1}$ and  $r_{2}$, respectively, containing an even number and an odd number of $1's$ of length $(N-2)$. For instance, the logical states for N =4 read
\begin{eqnarray}\label{logical states1}
|00\rangle_{L}=\frac{1}{\sqrt{2}}(|0000\rangle+|1111\rangle),\nonumber \\
|11\rangle_{L}=\frac{1}{\sqrt{2}}(|0110\rangle+|1001\rangle),\nonumber \\
|01\rangle_{L}=\frac{1}{\sqrt{2}}(|1010\rangle+|0101\rangle),\nonumber \\
|10\rangle_{L}=\frac{1}{\sqrt{2}}(|1100\rangle+|0011\rangle).
\end{eqnarray}

To implement two noncommuting holonomic single-logical-qubit gates and one nontrivial holonomic two-logical-qubit gate, one needs a set of operators to achieve the appropriate
transitions so that the evolution stays within the DFS. To this end, we need to seek for the operators that commute with the decoupling group $\mathcal{G}$. Here we consider the operators $\{\sigma_{1}^{x}\sigma_{j^{'}+1}^{x},\sigma_{j^{'}+1}^{z}\sigma_{N}^{z}\}$ ($j^{'}=1,2,\cdots,N-2$) which commute with the decoupling group $\mathcal{G}$. One can use a combination of the above operators to construct desired Hamiltonians, and as a result the DFS will not be destroyed.

\vspace{8pt}

\noindent{\it One qubit gates}.--
Explicitly, the forms of the Hamiltonians which generate a holonomic
single-qubit gate can be taken as follows
\begin{eqnarray}\label{H1}
H_{1}(t)&=&J_{1}(t)\sigma_{j+1}^{z}\sigma_{N}^{z},\nonumber\\
H_{1}^{'}(t)&=&J_{1}^{'}(t)(\cos\theta\sigma_{j+1}^{z}\sigma_{N}^{z}+\sin\theta\sigma_{1}^{x}\sigma_{j+1}^{x}),
\end{eqnarray}
where $J_{1}(t)$ and $J_{1}^{'}(t)$ are the controllable coupling parameters, $\theta$ is an arbitrary parameter,
and $j=1,\ldots,N-2$. The final time evolution operator which is composed by two-step evolutions reads
$U_{1}(T_{1},0)=\exp(-i\int_{\tau_{1}}^{T_{1}}H_{1}^{'}(t)dt)\exp(-i\int_{0}^{\tau_{1}}H_{1}(t)dt)$, 
where $\tau_{1}$ is an intermediate time parameter and $T_{1}$ is the evolution period.
Adjust the parameters such that
$\int_{0}^{\tau_{1}}J_{1}(t)dt=\int_{\tau_{1}}^{T_{1}}J_{1}^{'}(t)dt=\frac{\pi}{2}$,
we show that the evolution leads to a single-logical-qubit gate. Take $N=4$ and $j=1$ as an example, we have the evolution operator act on the logical
states in the DFS (\ref{logical states1}),
\begin{eqnarray}
&&U_{1}(T_{1},0)|00\rangle_{L}=-(\cos\theta|0\rangle_{L}+\sin\theta|1\rangle_{L})\otimes|0\rangle_{L}, \nonumber\\
&&U_{1}(T_{1},0)|01\rangle_{L}=-(\cos\theta|0\rangle_{L}+\sin\theta|1\rangle_{L})\otimes|1\rangle_{L}, \nonumber\\
&&U_{1}(T_{1},0)|10\rangle_{L}=-(-\sin\theta|0\rangle_{L}+\cos\theta|1\rangle_{L})\otimes|0\rangle_{L},\nonumber\\
&&U_{1}(T_{1},0)|11\rangle_{L}=-(-\sin\theta|0\rangle_{L}+\cos\theta|1\rangle_{L})\otimes|1\rangle_{L}.
\end{eqnarray}
It is clear that the resulting unitary operator can be written
in the subspace spanned by (\ref{logical states1}) by ignoring global phase as follows
$U_{1}(T_{1},0)=e^{-i\theta Y^{(1)}_{L}}\otimes I^{(2)}$.
where $Y^{(1)}_{L}=-i|0\rangle^{(1)}_{L}\langle1|^{(1)}_{L}+i|1\rangle^{(1)}_{L}\langle0|^{(1)}_{L}$ is the Pauli $Y$ operator acting on the  $1$-th logical qubit and $I^{(2)}$ is the identity matrix acting on the  $2$-th logical qubit.
It is straightforward to obtain the evolution operator in the subspace spanned by $(N-2)$ logical states (\ref{logical states}) up to a global phase as
\begin{eqnarray}\label{U11}
U_{1}(T_{1},0)=I^{(1)}\otimes\cdots\otimes e^{-i\theta  Y^{(j)}_{L}} \otimes\cdots \otimes I^{(N-2)},
\end{eqnarray}
where $N$ and $j$ are arbitrary, $Y^{(j)}_{L}=-i|0\rangle^{(j)}_{L}\langle1|^{(j)}_{L}+i|1\rangle^{(j)}_{L}\langle0|^{(j)}_{L}$ is the Pauli $Y$ operator acting on the  $j$-th logical qubit. This operator is nothing but one $j$-th single-logical-qubit  gate  ($j=1,\ldots,N-2$). It is shown that the unitary operator $U_{1}(T_{1},0)$ is purely holonomic according to the conditions of non-adiabatic HQC (see Methods).

We next explore the realization of another holonomic $j$-th single-logical-qubit  gate  ($j=1,\ldots,N-2$).
The desired Hamiltonians read
\begin{eqnarray}\label{H2}
H_{2}(t)&=&J_{2}(t)\sigma_{1}^{x}\sigma_{j+1}^{x},\nonumber\\
H_{2}^{'}(t)&=&J_{2}^{'}(t)\sigma_{1}^{x}\sigma_{j+1}^{x}, \end{eqnarray}
where $J_{2}(t)$ and $J_{2}^{'}(t)$ are the controllable coupling parameters.
With the two Hamiltonians and the $H_{1}(t)$ and $H_{1}^{'}(t)$ in Eq. (\ref{H1}),
the evolution operator which is composed by four-step evolution is given by
$U_{2}(T_{2},0)=\exp(-i\int_{\tau_{2}^{''}}^{T_{2}}H_{2}^{'}(t)dt)\exp(-i\int_{\tau_{2}^{'}}^{\tau_{2}^{''}}H_{1}^{'}(t)dt) \exp(-i\int_{\tau_{2}}^{\tau_{2}^{'}}H_{1}(t)dt)\exp(-i\int_{0}^{\tau_{2}}H_{2}(t)dt)$.
In the above equation, $\tau_{2}$, $\tau_{2}^{'}$, $\tau_{2}^{''}$ and $T_{2}$ are respectively intermediate time parameters and the evolution period. By choosing the following conditions
$\int_{0}^{\tau_{2}}J_{2}(t)dt=-\frac{\pi}{4}$, 
$\int_{\tau_{2}}^{\tau_{2}^{'}}J_{1}(t)dt=\int_{\tau_{2}^{'}}^{\tau_{2}^{''}}J_{1}^{'}(t)dt=\frac{\pi}{2}$, 
$\int_{\tau_{2}^{''}}^{T_{2}}J_{2}^{'}(t)dt=\frac{\pi}{4}$,
and the action of the unitary evolution operator $U_{2}(T_{2},0)$ is obtained for $N=4$ and $j=1$,
\begin{eqnarray}
&&U_{2}(T_{2},0)|00\rangle_{L}=-e^{-i\theta}|00\rangle_{L}, \nonumber\\
&&U_{2}(T_{2},0)|01\rangle_{L}=-e^{-i\theta}|01\rangle_{L}, \nonumber\\
&&U_{2}(T_{2},0)|10\rangle_{L}=-e^{i\theta}|10\rangle_{L},\nonumber\\
&&U_{2}(T_{2},0)|11\rangle_{L}=-e^{i\theta}|11\rangle_{L}.
\end{eqnarray}
Up to a global phase, the resulting unitary operator is of the form,
$U_{2}(T_{2},0)=e^{-i\theta Z^{(1)}_{L}}\otimes I^{(2)}$.
where $Z^{(1)}_{L}=|0\rangle^{(1)}_{L}\langle0|^{(1)}_{L}-|1\rangle^{(1)}_{L}\langle1|^{(1)}_{L}$ is the logical Pauli $Z$ operator  acting on the  $1$-th logical qubit and $I^{(2)}$ is the identity matrix acting on the  $2$-th logical qubit.
For arbitrary $N$ and $j$, it is not difficult to find the evolution operator in the subspace spanned by $(N-2)$ logical states (\ref{logical states}) by neglecting global phase,
\begin{eqnarray}\label{U22}
U_{2}(T_{2},0)=I^{(1)}\otimes\cdots\otimes e^{-i\theta  Z^{(j)}_{L}} \otimes\cdots \otimes I^{(N-2)},
\end{eqnarray}
where $Z^{(j)}_{L}=|0\rangle^{(j)}_{L}\langle0|^{(j)}_{L}-|1\rangle^{(j)}_{L}\langle1|^{(j)}_{L}$ is the logical Pauli $Z$ operator  acting on the  $j$-th logical qubit. Therefore we get another $j$-th single-logical-qubit gate  ($j=1,\ldots,N-2$), which commutes with
$U_{1}(T_{1},0)$ in (\ref{U11}). Similar to the illustration of the geometric property of $U_{1}(T_{1},0)$,
one can verify that the unitary operator $U_{2}(T_{2},0)$ also possesses holonomic property (see Methods).

As well known is that any single-logical-qubit rotation can be realized by arbitrary rotations around two orthogonal axes. Thus the above two noncommutative single-logical-qubit gates  $U_{1}=e^{-i\theta Y^{(j)}_{L}}$  and $U_{2}=e^{-i\theta Z^{(j)}_{L}}$, can realize any single-logical-qubit rotation.

\vspace{8pt}

\noindent{\it Two qubit gate}.--
To achieve a universal set of quantum gates, we now demonstrate how to realize  an entangling gate between the $k$-th logical qubit and the $l$-th logical qubit ($k<l=2,\ldots,N-2$) in the DFS spanned by (\ref{logical states}) using the generalized off-diagonal geometric proposal \cite{Mousolou}.
The required Hamiltonians are
\begin{eqnarray}\label{H3}
H_{3}(t)&=&J_{3}(t)(\cos\phi\sigma_{1}^{x}\sigma_{k+1}^{x}-\sin\phi\sigma_{k+1}^{z}\sigma_{l+1}^{z}), \nonumber\\
H_{3}^{'}(t)&=&J_{3}^{'}(t)\sigma_{1}^{x}\sigma_{k+1}^{x},
\end{eqnarray}
where  $\phi$ is an arbitrary parameter, and  $J_{3}(t)$ and $J_{3}^{'}(t)$ are the  controllable coupling parameters.
The final time evolution operator resulted from the two-step evolution is
$U_{3}(T_{3},0)=\exp(-i\int_{\tau_{3}}^{T_{3}}H_{3}^{'}(t)dt)\exp(-i\int_{0}^{\tau_{3}}H_{3}(t)dt)$,
where $\tau_{3}$ and $T_{3}$ are respectively an intermediate time parameter and the evolution period.
Control the parameters to make sure that
$\int_{0}^{\tau_{3}}J_{3}(t)dt=\int_{\tau_{3}}^{T_{3}}J_{3}^{'}(t)dt=\frac{\pi}{2}$,
we have $U_{3}(T_{3},0)$ written in the DFS formed by (\ref{logical states1}) for $N=4$, $k=1$ and $l=2$,
\begin{eqnarray}
&&U_{3}(T_{3},0)|00\rangle_{L}=-(\cos\phi|00\rangle_{L}-\sin\phi|10\rangle_{L}), \nonumber\\
&&U_{3}(T_{3},0)|01\rangle_{L}=-(\cos\phi|01\rangle_{L}+\sin\phi|11\rangle_{L}), \nonumber\\
&&U_{3}(T_{3},0)|10\rangle_{L}=-(\sin\phi|00\rangle_{L}+\cos\phi|10\rangle_{L}),\nonumber\\
&&U_{3}(T_{3},0)|11\rangle_{L}=-(-\sin\phi|01\rangle_{L}+\cos\phi|11\rangle_{L}).
\end{eqnarray}
The unitary operator is of an equivalent form
$U_{3}(T_{3},0)=e^{i\phi Y^{(1)}_{L}\otimes Z^{(2)}_{L}}$ (up to global phase).
Furthermore, take $N=6$, $k=1$ and $l=2$, the action of $U_{3}(T_{3},0)$ on the logical states in the logic DFS (\ref{logical states}) can be found as
\begin{eqnarray}
&&U_{3}(T_{3},0)|00mn\rangle_{L}=-(\cos\phi|00\rangle_{L}-\sin\phi|10\rangle_{L})\otimes|mn\rangle, \nonumber\\
&&U_{3}(T_{3},0)|01mn\rangle_{L}=-(\cos\phi|01\rangle_{L}+\sin\phi|11\rangle_{L})\otimes|mn\rangle, \nonumber\\
&&U_{3}(T_{3},0)|10mn\rangle_{L}=-(\sin\phi|00\rangle_{L}+\cos\phi|10\rangle_{L})\otimes|mn\rangle,\nonumber\\
&&U_{3}(T_{3},0)|11mn\rangle_{L}=-(-\sin\phi|01\rangle_{L}+\cos\phi|11\rangle_{L})\otimes|mn\rangle,
\end{eqnarray}
where $m,n\in\{0,1\}$. The resulting unitary operator can be written
in the subspace spanned by (\ref{logical states}) as follows by ignoring global phase,
$U_{3}(T_{3},0)=e^{i\phi Y^{(1)}_{L}\otimes Z^{(2)}_{L}}\otimes I^{(3)}\otimes I^{(4)}$.
Meanwhile, for $N=6$, $k=2$ and $l=3$, we get
\begin{eqnarray}
&&U_{3}(T_{3},0)|m00n\rangle_{L}=-|m\rangle\otimes(\cos\phi|00\rangle_{L}-\sin\phi|10\rangle_{L})\otimes|n\rangle, \nonumber\\
&&U_{3}(T_{3},0)|m01n\rangle_{L}=-|m\rangle\otimes(\cos\phi|01\rangle_{L}+\sin\phi|11\rangle_{L})\otimes|n\rangle, \nonumber\\
&&U_{3}(T_{3},0)|m10n\rangle_{L}=-|m\rangle\otimes(\sin\phi|00\rangle_{L}+\cos\phi|10\rangle_{L})\otimes|n\rangle,\nonumber\\
&&U_{3}(T_{3},0)|m11n\rangle_{L}=-|m\rangle\otimes(-\sin\phi|01\rangle_{L}+\cos\phi|11\rangle_{L})\otimes|n\rangle,
\end{eqnarray}
where $m,n\in\{0,1\}$. In this case, the unitary operator is
$U_{3}(T_{3},0)=I^{(1)}\otimes e^{i\phi Y^{(2)}_{L}\otimes Z^{(3)}_{L}}\otimes I^{(4)}$ up to a global phase.
It is easy to generalize the results to any $N,k,j$ and the evolution operator reads
\begin{eqnarray}\label{U33}
U_{3}(T_{3},0)=I^{(1)}\otimes\cdots\otimes e^{i\phi Y^{(k)}_{L}\otimes Z^{(l)}_{L}}\otimes\cdots \otimes I^{(N-2)},
\end{eqnarray}
 in the subspace spanned by $(N-2)$ logical states (\ref{logical states}). One can find that $U_{3}(T_{3},0)$ is a nontrivial entangling logical gate when $\sin\phi$ and $\cos\phi$ are nonzero. The geometric feature of $U_{3}(T_{3},0)$ can be demonstrated by resorting to the eigenstates of $Y_L^{(1)}$ and $Y_L^{(2)}$ as we did for $U_{1}(T_{1},0)$ (see Methods). As a result, we have achieved a universal set of non-adiabatic holonomic quantum gates in DFSs of the decoupling group $\mathcal{G}$ with two non-commutative single-logical-qubit gates and one non-trivial holonomic two-qubit gate.

\vspace{8pt}
\noindent{\bf Discussions}

We next discuss the physical realization of our scheme in physical systems. The above-mentioned two-body qubit-qubit interactions required for the implementation of the quantum logic gates may be achieved in coupled microcavity system, and that is an array of cavities coupled via exchange of virtual photons with one $\Lambda$-type three-level atom in each cavity~\cite{microcavities}. In the literature, an anisotropic Heisenberg spin-1/2 lattice in an external magnetic field was proposed by individually adjusting the external lasers which were illuminated on the atoms. The effective Hamiltonian is of the form
\begin{eqnarray}
H_{\rm eff}=\sum_{i=1}^{N}\bigg(J'_{z}\sigma_i^z+J_x \sigma_i^x\sigma_{i+1}^x + J_y \sigma_i^y\sigma_{i+1}^y+J_z \sigma_i^z\sigma_{i+1}^z\bigg)
\label{Hp}
\end{eqnarray}
where the parameters $J'_{z}$, $J_{x,y,z}$ can individually be tuned via external lasers through controlling the laser frequencies, Rabi frequencies and the cavity-cavity couplings~\cite{microcavities}. Based on the results, different kinds of two-body qubit-qubit interactions can be generated by suitably selecting the parameters $J'_{z}$, $J_{x,y,z}$, so our proposed logic gates may be realized in the coupled microcavity system.
According to the effective qubit-qubit interaction, nearest neighbor couplings of qubits can be realized. Our desired $H_{1,2,3}$ and $H_{1,2,3}^{'}$ are based on two-qubit interactions including the cases that the two qubits are next to each other or not. The two-qubit interactions may be achievable in the coupled microcavities by controlling the couplings of different microcavities based on Hamiltonian (\ref{Hp}). We take $H_1$ as an example to explain the physical realization of the interaction. Number the atoms in each microcavity as $1$ to $N$. Let $j+1$-th and $N$-th microcavities interact with each other while the others do not. Adjust the detunings and Rabi frequencies in the two specified microcavities such that $J_x$ and $J_y$ are zero~\cite{microcavities}, we get $H_1$. The other target two-qubit interactions can be obtained similarly.

In this work, we have explored the implementation of universal sets of non-adiabatic holonomic quantum gates by considering a linear system-bath interaction Hamiltonian in which each qubit has its own bath. The holonomic quantm gates are achieved in the DFSs of the decoupling group. Our results possess four-fold merits. Firstly, the quantum operations bear non-adiabatic holonomic property and hence they are robust against control imprecisions and require less operation time. Secondly, based on combination of the active DD and the passive DFSs, the quantum operations are resisted to the decoherence caused by unavoidable interaction with environment. Thirdly, our scheme is realizable by utilizing only two-body interactions rather than many-body interactions. From the perspective of experiments, two-body interactions are easier to achieve in physical systems than many-body interactions. Lastly, our encoding strategy with an encoding rate of $\frac{N-2}{N}$ makes our scheme preferable consider the scalability of quantum computation to many logical qubits. In the following we would like to compare our work with the one presented in Ref.~\cite{Long14} in which non-adiabatic HQC was also proposed in the DFS by DD based on two-qubit interactions. Compared with Ref.~\cite{Long14}, our scheme exhibits two desirable advantages. One is about the encoding rate, it is $\frac{N-2}{N}$ in our scheme, while in Ref.~\cite{Long14} it is $\frac{1}{3}$. The increased encoding rate is due to the fact that we encode our logical qubits in the DFS provided by the dynamical decoupling itself and hence our encoding structure is more symmetric. The other advantage is that, in our scheme any arbitrary single-logical-qubit gate can be obtained by simple combinations of the two single-logical-qubit gates proposed, where it is not the case in Ref.~\cite{Long14}.
Therefore our results reduce the complexity of practical implementation of holonomic quantum gates in the DFSs of the decoupling group. We expect our scheme can shed light on the experimentally achievable implementations of HQC in DFSs.

\vspace{8pt}
\noindent{\bf Methods}

We need to verify whether the unitary operators $U_{1,2,3}$ are purely holonomic quantum gates.
The conditions of non-adiabatic HQC has been proposed in Refs. \cite{Tong,Xu}. Consider an $N$-dimensional quantum system with Hamiltonian $H_{S}(t)$. Assume there exists a time-dependent $K$-dimensional subspace $\mathcal{M}(t)$ spanned by a set of orthonormal basis vectors $\{|\Psi_{k}(t)\rangle,k=1,\ldots,K\}$ at each time $t$. Here $|\Psi_{k}(t)\rangle$ can be obtained from the Schr$\ddot{o}$dinger equation
$|\Psi_{k}(t)\rangle=\mathbf{T}\exp(-i\int_{0}^{t}H_{S}(t')dt')|\Psi_{k}(0)\rangle=U(t,0)|\Psi_{k}(0)\rangle$, 
with $k=1,\ldots,K$, and $\mathbf{T}$ is the time ordering operator. The unitary transformation $U(\tau,0)=\mathbf{T}\exp(-i\int_{0}^{\tau}H(t')dt')$ is a holonomy matrix acting on the subspace $\mathcal{M}(0)$ if $\{|\Psi_{k}(t)\rangle\}$ satisfy the two conditions:
$(i) \sum_{k=1}^{K}|\Psi_{k}(\tau)\rangle\langle\Psi_{k}(\tau)|=\sum_{k=1}^{K}|\Psi_{k}(0)\rangle\langle\Psi_{k}(0)|$, 
 and $(ii) \langle\Psi_{k}(t)|H(t)|\Psi_{l}(t)\rangle=0, ~~~ k,l=1,\ldots,K$,
where $\tau$ is the evolution period. Condition $(i)$ ensures that the states in the subspace $\mathcal{M}(0)$ complete a cyclic evolution, and condition $(ii)$ ensures that the cyclic evolution is purely geometric.
\vspace{8pt}

\begin{figure}[tbp]
\includegraphics[width=100mm]{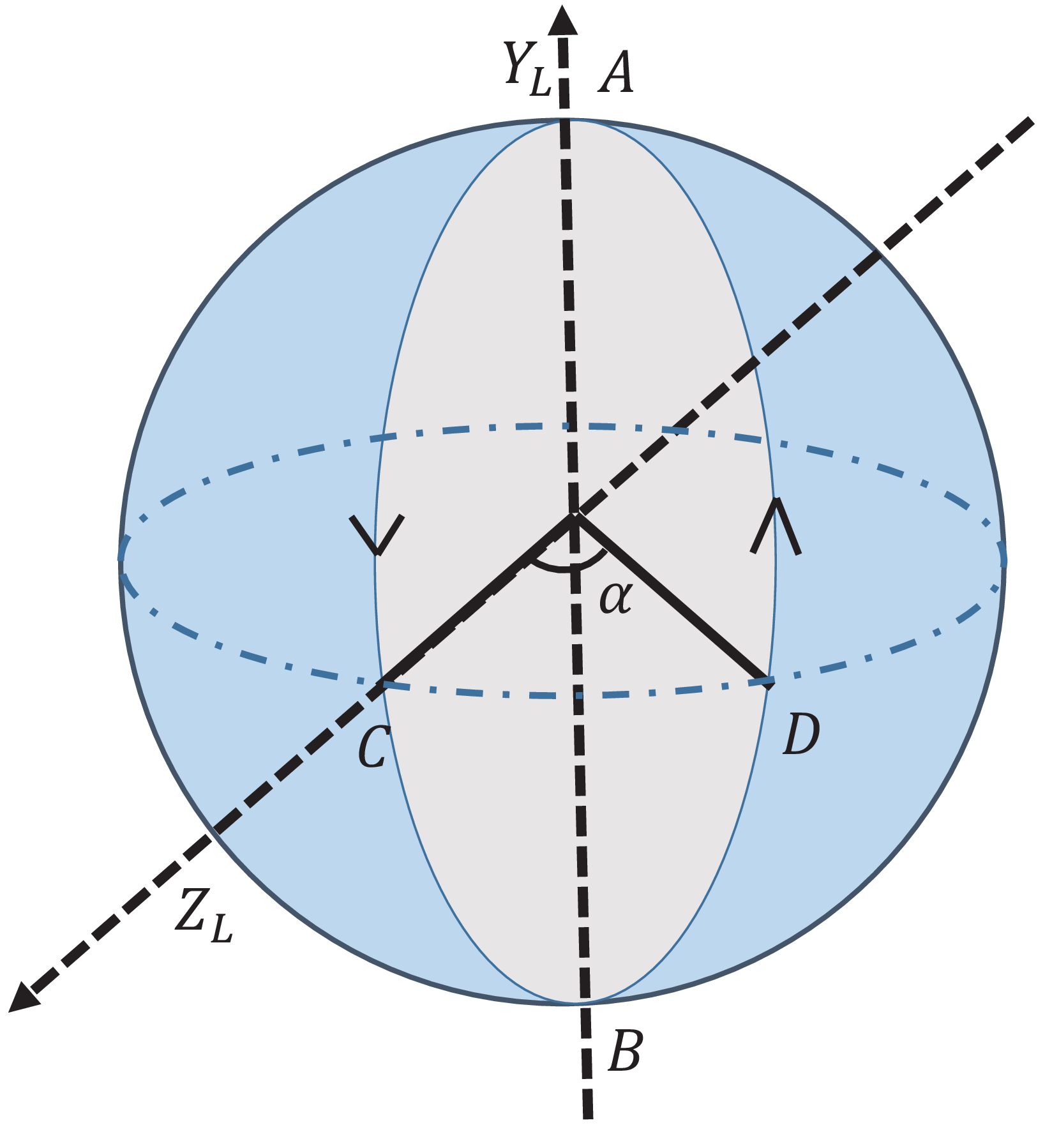}\\
\caption{Illustration of geometric property of two noncommuting single-logical-qubit gates $U_{1}(T_{1},0)$ and $U_{2}(T_{2},0)$ in logical Bloch sphere. }
\label{fig1}
\end{figure}

\noindent{\it Holonomic property of $U_{1}$}.--
Here we explore the holonimic property of $U_{1}$ by an example with $N=4$ and $j=1$ by considering the orthonormal basis vectors $\{|\Psi_{1}(0)\rangle=\frac{1}{\sqrt{2}}(|0\rangle_{L}+i|1\rangle_{L})\otimes|0\rangle_{L},
|\Psi_{2}(0)\rangle=\frac{1}{\sqrt{2}}(|0\rangle_{L}-i|1\rangle_{L})\otimes|0\rangle_{L},|\Psi_{3}(0)\rangle=\frac{1}{\sqrt{2}}(|0\rangle_{L}+i|1\rangle_{L})\otimes|1\rangle_{L},
|\Psi_{4}(0)\rangle=\frac{1}{\sqrt{2}}(|0\rangle_{L}-i|1\rangle_{L})\otimes|1\rangle_{L}\}$.
Condition (i) is satisfied since the subspace spanned by
$\{U_{1}(T_{1},0)|\Psi_{k}(0)\rangle\}$ coincides with $\{|\Psi_{k}(0)\rangle,k=1,2,3,4\}$. Condition (ii) needs $\langle\Psi_{k}(0)|U(t,0)^{\dagger} H(t) U(t,0)|\Psi_{l}(0)\rangle=0$.
This condition can be written as $\langle\Psi_{k}(0)| H_{1}(t)|\Psi_{l}(0)\rangle=0$ and $\langle\Psi_{k}(\tau_{1})| H^{'}_{1}(t)|\Psi_{l}(\tau_{1})\rangle=0$
because $H_{1}(t)$ and $H^{'}_{1}(t)$ respectively commute
with their evolution operators. It is easy to see that $\langle\Psi_{k}(0)| H_{1}(t)|\Psi_{l}(0)\rangle=0$ and $\langle\Psi_{k}(\tau_{1})| H^{'}_{1}(t)|\Psi_{l}(\tau_{1})\rangle=0$. Thus, both conditions (i) and (ii) are satisfied, and $U_{1}(T_{1},0)$ is a holonomic single-logical-qubit gate. One can also illustrate the geometric property of $U_{1}(T_{1},0)$ by visualizing the evolution in logical Bloch sphere as shown in Fig. \ref{fig1}. The Hamiltonians $H_{1}(t)$ and $H^{'}_{1}(t)$ drive the eigenstates of $Y^{(j)}_{L}$ from point A  with the eigenvalue $+1$
to the opposite pole B  with the eigenvalue $-1$ and then back to point A, which  completes a loop along the geodesic line ACBDA.
Therefore there is no dynamical contribution during the whole evolution and
 the single-logical-qubit gate $U_{1}(T_{1},0)$ is purely geometric.


\noindent{\it Holonomic property of $U_{2}$}.--
We look at the example with $N=4$ and $j=1$ again for the demonstration of the holonomic property of $U_{2}$, and consider the orthonormal basis vectors $\{|\Psi_{1}(0)\rangle=|00\rangle_{L},
|\Psi_{2}(0)\rangle=|01\rangle_{L},|\Psi_{3}(0)\rangle=|10\rangle_{L},
|\Psi_{4}(0)\rangle=|11\rangle_{L}\}$.
Condition (i) is fulfilled since the subspace spanned by
$\{U_{2}(T_{2},0)|\Psi_{k}(0)\rangle\}$ coincides with $\{|\Psi_{k}(0)\rangle,k=1,2,3,4\}$.
Furthermore,  one needs to verify that condition (ii) is satisfied, i.e., $\langle\Psi_{k}(0)|U(t,0)^{\dagger} H(t) U(t,0)|\Psi_{l}(0)\rangle=0$.
The condition can be rewritten as $\langle\Psi_{k}(0)| H_{2}(t)|\Psi_{l}(0)\rangle=0$, $\langle\Psi_{k}(\tau_{2})| H_{1}(t)|\Psi_{l}(\tau_{2})\rangle=0$, $\langle\Psi_{k}(\tau^{'}_{2})| H^{'}_{1}(t)|\Psi_{l}(\tau^{'}_{2})\rangle=0$ and $\langle\Psi_{k}(\tau^{''}_{2})| H^{'}_{2}(t)|\Psi_{l}(\tau^{''}_{2})\rangle=0$
because $H_{2}(t)$,  $H_{1}(t)$, $H^{'}_{1}(t)$ and   $H^{'}_{2}(t)$  respectively commute
with their evolution operators.  We thus find Conditions (ii) is satisfied as well, and therefore $U_{2}(T_{2},0)$ is  a holonomic single-logical-qubit gate. Similarly, one can also illustrate the geometric property of $U_{2}(T_{2},0)$ by Fig. \ref{fig1}. The Hamiltonians $H_{2}(t)$,  $H_{1}(t)$, $H^{'}_{1}(t)$ and   $H^{'}_{2}(t)$ drive the eigenstates of $Z^{(j)}_{L}$ from point C  with the eigenvalue $+1$  completes a  cyclic evolution along the geodesic line CBDAC.
Hence the single-logical-qubit gate $U_{2}(T_{2},0)$ is purely geometric.
\vspace{8pt}

\noindent{\it Holonomic property of $U_{3}$}.--
We take $N=4$, $k=1$ and $l=2$ as an example to show the holonomic property of $U_{3}$. By defining $\ket{\bar{0}}_{L}=\frac{1}{\sqrt{2}}(\ket{0}_{L}+i\ket{1}_{L})$ and $\ket{\bar{1}}_{L}=\frac{1}{\sqrt{2}}(\ket{0}_{L}-i\ket{1}_{L})$, the two logical qubit states $\{\ket{\bar{0}\bar{0}}_{L},\ket{\bar{0}\bar{1}}_{L},\ket{\bar{1}\bar{0}}_{L},\ket{\bar{1}\bar{1}}_{L}\}$ form a basis of the 4 dimensional Hilbert space $\mathcal{S}$. Now we split $\mathcal{S}$ into two orthogonal subspaces $\mathcal{S}_1=$Span$\{\ket{\bar{0}\bar{0}}_{L},\ket{\bar{0}\bar{1}}_{L}\}$ and $\mathcal{S}_2=$Span$\{\ket{\bar{1}\bar{0}}_{L},\ket{\bar{1}\bar{1}}_{L}\}$, and in the representation the Hamiltonian $H_3(t)$ and $H_3^{\prime}(t)$ read
$H_3(t)=J_3(t)\left(
               \begin{array}{cc}
                 0 & A \\
                 A^{\dag} & 0 \\
               \end{array}
             \right)$,
$H_3^{\prime}(t)=J_3^{\prime}(t)\left(
               \begin{array}{cc}
                 0 & B \\
                 B^{\dag} & 0 \\
               \end{array}
             \right)$,
where the  matrices $A$ and $B$ can be written as $\left(
                                                \begin{array}{cc}
                                                  -i\cos\phi & -\sin\phi \\
                                                  -\sin\phi & -i\cos\phi\\
                                                \end{array}
                                              \right)
$ and $\left(
        \begin{array}{cc}
          -i & 0 \\
          0 & -i \\
        \end{array}
      \right)$, respectively.
The corresponding evolution operators for the two steps read
\begin{equation}
U_3(\tau_{3},0)=-i\left(
          \begin{array}{cc}
            0 & A \\
            A^{\dag} & 0 \\
          \end{array}
        \right),
U_3(T_{3},\tau_{3})=-i\left(
          \begin{array}{cc}
            0 & B \\
            B^{\dag} & 0 \\
          \end{array}
        \right),
\end{equation}respectively and $U_3(T_{3},0)$ can be shown as
\begin{equation}
U_3(T_{3},0)=-\left(
       \begin{array}{cc}
         BA^{\dag} & 0 \\
         0 & B^{\dag}A \\
       \end{array}
     \right).
\end{equation}
The underlying idea is that, at time $\tau_3$, the two subspaces $\{\mathcal{S}_1,\mathcal{S}_2\}$ evolved into each other and then, at time $T_3$, they return, and this leads to a loop evolution in the Hilbert space and therefore condition (i) is satisfied.
We then check that condition (ii) is satisfied, i.e., $\langle\Psi_{k}(0)|U(t,0)^{\dagger} H(t) U(t,0)|\Psi_{l}(0)\rangle=0$. This is equivalent to $\langle\Psi_{k}(0)| H_{3}(t)|\Psi_{l}(0)\rangle=0$ and $\langle\Psi_{k}(\tau_{3})| H^{'}_{3}(t)|\Psi_{l}(\tau_{3})\rangle=0$ because $H_{3}(t)$ and $H^{'}_{3}(t)$ respectively commute with their evolution operators. Thus, both conditions (i) and (ii) are satisfied, and $U_{3}(T_{3},0)$ is a holonomic two-logical-qubit gate.

The set of a $2$-dimensional subspaces $\{\mathcal{S}_1,\mathcal{S}_2\}$ in the $4$-dimensional Hilbert space forms a Grassman manifold $G(4;2)$. The closed path $\mathbf{C}$ of $2$-dimensional subspaces is a loop in $G(4; 2)$. The set of all bases forms a Stiefel manifold $\mathcal{S}(4; 2)$, which is a fiber bundle with $G(4; 2)$ as base manifold and with the set of $2\times 2$ unitary matrices as fibers. The two steps of evolution to achieve $U_{3}(T_{3},0)$ correspond to two geodesic lines in $G(4;2)$, that transform $\mathcal{S}_1=$Span$\{\ket{\bar{0}\bar{0}}_{L},\ket{\bar{0}\bar{1}}_{L}\}$ to its orthogonal complement $\mathcal{S}_2=$Span$\{\ket{\bar{1}\bar{0}}_{L},\ket{\bar{1}\bar{1}}_{L}\}$  and then back to $\mathcal{S}_1=$Span$\{\ket{\bar{0}\bar{0}}_{L},\ket{\bar{0}\bar{1}}_{L}\}$ along the geodesic lines. The accompanying non-Abelian geometric phase represents the $2\times2$ fiber on the base manifold of $\mathcal{S}(4; 2)$.

\vspace{8pt}
\noindent{\it Performance of the quantum gates with imperfect DD sequences}.--
The fact that our holonomic quantum gates are resistant to non-collective decoherence is based on the DD approach. As a result, the existence of DD pulse errors will affect the performance of our proposed quantum gates. Here we provide some numerical results to demonstrate the effects of DD pulse errors.
The decoupling strategy utilized in our work can be described by alternatively applying computational and DD operations with $XY-4$ sequence as the basic DD sequence.

\begin{figure}[tbp]
\includegraphics[width=90mm]{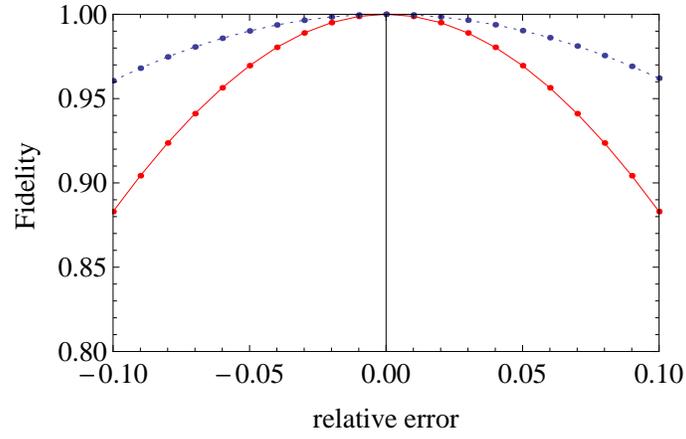}\\
\caption{Numerical results of the fidelity of the two-qubit logical gate
$e^{-i\frac{\pi}{4}Y_{L}^{(1)}\otimes Z_{L}^{(2)}}$ in the presence of the flip-angle error (Red solid curve) and frequency detuning error (blue dotted curve). The parameters are chosen as follows, $-0.1\leq\epsilon\leq0.1$ and $-0.1\leq\delta\leq0.1$.}
\label{fig2}
\end{figure}

One of the main errors in DD sequences is flip-angle error caused by the inaccuracy in pulse duration and Rabi frequency.
With a relative flip-angle error $\epsilon$, the imperfect pulse propagator reads ~\cite{Long14}
\begin{eqnarray}
R_{f}(\vartheta_{p})=e^{-i \sigma^{\alpha}_{i}(1+\epsilon)\vartheta_{p}/2},
\label{Rf}
\end{eqnarray}
where $f$ indicates the effect of the flip-angle error, $\sigma_{i}^{\alpha}$ ($\alpha=x,y,z$) are Pauli matrices acting on the $i$-th physical qubit and $\vartheta_{p}$ is the rotation angle about the $\alpha$ axis. The angle $\vartheta_{p}$ is $\pi$ for ideal instantaneous pulses. 
The fidelity of the quantum gates can be computed numerically according to the following formula~\cite{Long14},
\begin{eqnarray}
F=\frac{|Tr(U_{id} U_{im}^{\dagger})|}{\sqrt{Tr(U_{id} U_{id}^{\dagger})Tr(U_{im} U_{im}^{\dagger})}},
\label{F}
\end{eqnarray}
where $U_{id}(U_{im})$ is the ideal (imperfect) propagator without (with) DD pulse errors. We take the two-logical-qubit holonomic gate as an example to show the performance of our scheme in the presence of the flip-angle error. As shown in Fig. \ref{fig2}, it is clear that the type of error destroys the gate fidelity severely when $|\epsilon|>0.02$ (see the red solid curve).

Another main error source in DD sequences in due to the frequency detuning error which usually leads to errors in the rotation angle and the direction of the rotation axis. With a relative detuning error $\delta$, the imperfect rotation operator is of the form~\cite{Long14},
\begin{eqnarray}
R_{d}(\vartheta_{p})=\cos(\frac{\vartheta_{p}\sqrt{1+\delta^{2}}}{2})I-i\sin(\frac{\vartheta_{p}\sqrt{1+\delta^{2}}}{2}) \vec{\sigma}.\vec{n}_{d},
\label{Rd}
\end{eqnarray}
where $d$ indicates the effect of frequency detuning error, and the actual rotation axis is $\vec{n}_{d}=(\cos \varphi/\sqrt{1+\delta^{2}}, \sin \varphi/\sqrt{1+\delta^{2}}, d/\sqrt{1+\delta^{2}})$. According to  Eq. (\ref{Rd}), we numerically find the fidelity of the two-logical-qubit holonomic gate when the frequency detuning error is presented (see Fig. \ref{fig2}, blue dotted curve). Our results show that the quantum gate is more tolerant to the detuning error than the flip-angle error, and the results are consistent with those given in Ref.~\cite{Long14}. Hence our scheme requires high precision in adjusting pulse duration and Rabi frequency in experiments.


\vspace{8pt}
{\bf Supplementary Information} is linked to the online version of the paper at www.nature.com/nature.
\vspace{2pt}

{\bf Acknowledgements}

This work was supported by the NSF of China (Grant Nos. 11205028, 11175043, 11405026 and 11405008), the Plan for Scientific and Technological Development of Jilin Province (Nos. 20130522145JH and 20150520083JH) and the
Fundamental Research Funds for the Central Universities (Grant Nos. 14QNJJ008 and 2412015KJ009). C.F.S. was also supported in part by the Government of China through CSC. X.L.F. is supported by the NSF of Shanghai (Grant No. 15ZR1430600). J.L.C. is supported by National Basic Research Program (973 Program) of China (Grant No. 2012CB921900) and the NSF of China (Grant Nos. 11175089 and 11475089).

\vspace{2pt}

{\bf Author contributions}

C.S. initiated the idea. All authors developed the scheme and wrote the main manuscript text.

\vspace{2pt}

{\bf Additional information}

Competing financial interests: The authors declare no competing financial interests.

Correspondence and requests for materials should be addressed to C. S. (suncf997@nenu.edu.cn) or G. W. (wanggc887@nenu.edu.cn) or C. W. (chunfeng\_wu@sutd.edu.sg).

\newpage

\end{document}